\PassOptionsToPackage{usenames,dvipsnames}{xcolor}
\documentclass{sig-alternate}

\usepackage{eurosym}
\usepackage{framed}
\usepackage{xspace}

\usepackage{graphicx}
\usepackage{caption}
\usepackage{subcaption}
\usepackage{alltt}
\usepackage{fancyvrb}

\usepackage{color}
\usepackage[usenames,dvipsnames]{xcolor}

\usepackage{balance}

\makeatletter
\def\@copyrightspace{\relax}
\makeatother

\def\pprw{8.5in}
\def\pprh{11in}

\definecolor{light-gray}{gray}{0.95}
\definecolor{mid-gray}{gray}{0.85}
\definecolor{darkred}{rgb}{0.7,0.25,0.25}
\definecolor{darkgreen}{rgb}{0.15,0.55,0.15}
\definecolor{darkblue}{rgb}{0.1,0.1,0.5}
\definecolor{blue}{rgb}{0.19,0.58,1}

\newtheorem{definition}{Definition}

\newcommand{\stitle}[1]{\noindent\textbf{#1}}
\newcommand{\sys}{Precision Interfaces\xspace}
\newcommand{\lang}{PILang\xspace}

\begin{document}

\title{Precision Interfaces}

\numberofauthors{3} 
\author{
\alignauthor
Haoci Zhang\\
       \affaddr{Tsinghua University}\\
       \email{zhanghaoci@gmail.com}
\alignauthor
Thibault Sellam\\
       \affaddr{Columbia University}\\
       \email{tsellam@cs.columbia.edu}
\alignauthor
Eugene Wu\\
       \affaddr{Columbia University}\\
       \email{ewu@cs.columbia.edu}
}

\maketitle

\begin{abstract}
Building interactive tools to support data analysis is hard because it is not always clear what to build and how to build it. To address this problem, we present Precision Interfaces, a semi-automatic system to generate task-specific data analytics interfaces. Precision Interface can turn a log of executed programs into an interface, by identifying micro-variations between the programs and mapping them to interface components. This paper focuses on SQL query logs, but we can generalize the approach to other languages. Our system operates in two steps: it first build an interaction graph, which describes how the queries can be transformed into each other. Then, it finds a set of UI components that covers a maximal number of transformations. To restrict the domain of changes to be detected, our system uses a domain-specific language, PILang. We give a full description of Precision Interface's components, showcase an early prototype on real program logs and discuss future research opportunities.
\end{abstract}

\section{Introduction}
\label{sec:intro}

Data analysis and exploration tools let users navigate their datasets through interface components such as dropdown lists, sliders, or buttons. Those applications dramatically accelerate analysis by abstracting out a \emph{common} set of operations and lifting them into the visual domain~\cite{schneiderman1986eight}. A successful application of this principle is Tableau, which optimizes OLAP exploration~\cite{stolte2002polaris}. Another example is Crossfilter, which targets filtering~\cite{squarecrossfilter}. Yet, building those applications is hard. The process involves high {\it development costs} in terms of time or expertise, combined with the reality that it is not always clear {\it what to build}. Therefore, interfaces do not exist for all but the most common and highest profile analysis tasks.


One approach is to provide tools and libraries that make it easier, perhaps for even end-users, to build interfaces. This is the rationale behind Shiny, a framework that helps statisticians quickly create Web interfaces for R scripts. Similarly, tools such as Sikuli~\cite{yeh2009sikuli} or Microsoft Access enable users with no engineering background to build software. Although easier to use than lower level libraries such as NodeJS or Bootstrap, they still require learning and practice. To illustrate, Shiny's Website claims that ``no HTML, CSS or JavaScript knowledge [is] required'', but its users need to understand reactive programming.  Simply put, programming is hard~\cite{evans1989best}.

When task-specific interfaces are not available, users default to more generic systems. For example, Tableau is a powerful interface for performing OLAP-based exploration, however any given task only utilizes a small fraction of the interface's capabilities (Section~\ref{sec:experiments} describes a case study in more detail).  In practice, users will use whatever tools are available on-hand, or rely on technical experts to perform the analysis on their behalf.
This approach has two important limits. It is not \emph{discoverable}~\cite{spool2005makes}: users often struggle to identify the features that will let them perform their analysis. It is not \emph{efficient}: the short cuts that could be tailored to a common task are do not exist in more general systems.
Ideally, users should have interfaces tailored to their set of tasks~\cite{gajos2010automatically, kim1993providing, myers2000past, weld2003automatically, zanden1990automatic}.
The programming languages community has seen this pattern in the rise of {\it domain specific languages}~\cite{heer2015predictive} and this can be viewed as the analogy for visual interfaces.

To this end we argue for \emph{Precision Interfaces}, an automatic tool to generate task-specific data analytics interfaces. We believe that two observations point towards the promise of Precision Interfaces.  First, modern applications~\cite{alspaugh2014better,alspaugh2014analyzing} and analysis frameworks~\cite{ground} are increasing storing rich metadata about user analysis operations, including the programs that are run; simply consider the query logs that nearly all databases maintain.   These traces indirectly capture the user's analytic needs and may be mined to identify patterns and common analyses that can be translated into interfaces.
Second, data analysis is inherently incremental~\cite{bates1989design}. Consequently, the programs in the log also change incrementally. By identifying these incremental changes, we can more readily map them to interface components.
With the confluence of these observations, we hope to move towards a future where {\it ``no interface is left behind''}.

\begin{figure}
    \centering
    \begin{subfigure}[b]{0.45\textwidth}
    	\begin{footnotesize}
    	\begin{Verbatim}[commandchars=\\\{\},frame=single]
Q1: SELECT * FROM Sales WHERE Country = \textcolor{Red}{'US'}
Q2: SELECT * FROM Sales WHERE Country = \textcolor{Red}{'UK'}

Q3: SELECT \textcolor{Red}{TOP 5} * FROM Sales
Q4: SELECT * FROM Sales
		\end{Verbatim}
		\end{footnotesize}
        \caption{Two pairs of consecutive queries.}
        \label{fig:query-pairs}
    \end{subfigure}
    \par\bigskip
    \begin{subfigure}[b]{0.45\textwidth}
    	\centering
        \includegraphics[width=.5\textwidth]{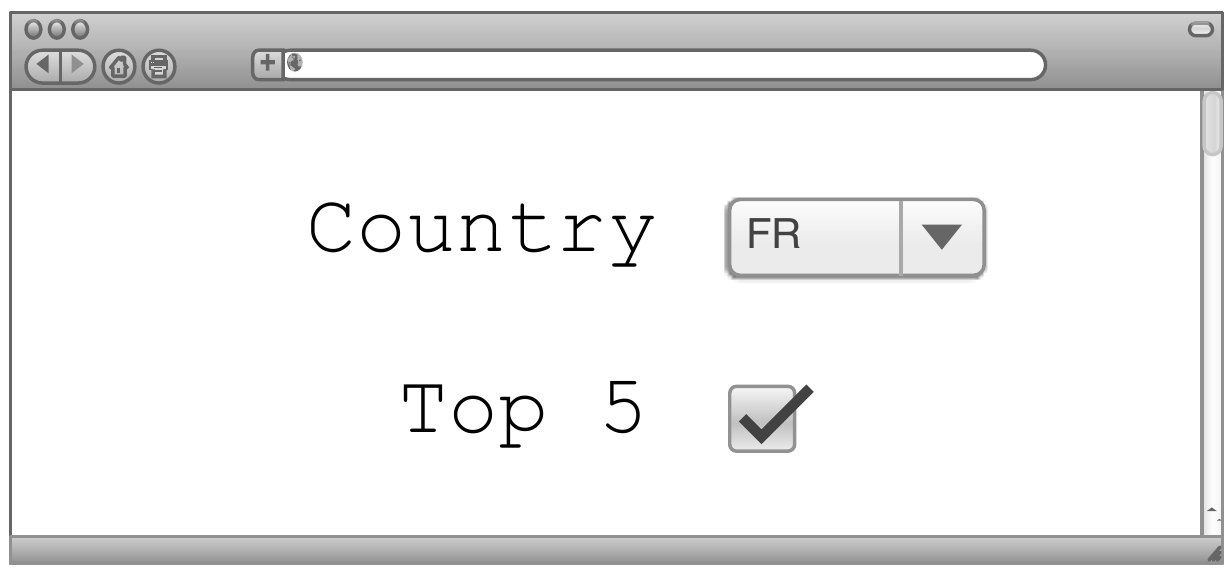}
        \caption{Matching interface.}
        \label{fig:little-interface}
    \end{subfigure}
    \caption{Example of automated interface design.}
    \label{fig:intro-example}
\end{figure}
In the rest of this paper, we will describe how to build data analytics interfaces from program logs. Our current prototype and examples focus on \emph{SQL query logs}, however the techniques can apply to any other language. The main idea is to detect small differences between programs, and map those to user interactions.  Consider for instance the two pairs of SQL queries pictured in Figure~\ref{fig:query-pairs}. We can transform Q1 into Q2 by changing the equality predicate in the \texttt{WHERE} clause. Similarly we can obtain Q4 from Q3 by adding a \texttt{TOP 5} statement. Precision Interfaces can recognize those interaction patterns, and infer an interface from them as shown in Figure~\ref{fig:little-interface}.

The rest of this paper is organized as follows. In Section~\ref{sec:overview}, we give an overview of the PI pipeline. Sections~\ref{sec:language} and ~\ref{sec:mapping} focus on two specific aspects of our system: how to describe interactions, and how to map widgets to interactions. We present a use case with real data in Section~\ref{sec:experiments}. We conclude in Section~\ref{sec:conclusion}.

\section{System Overview}
\label{sec:overview}

\begin{figure}
	\centering
	\includegraphics[width=\columnwidth]{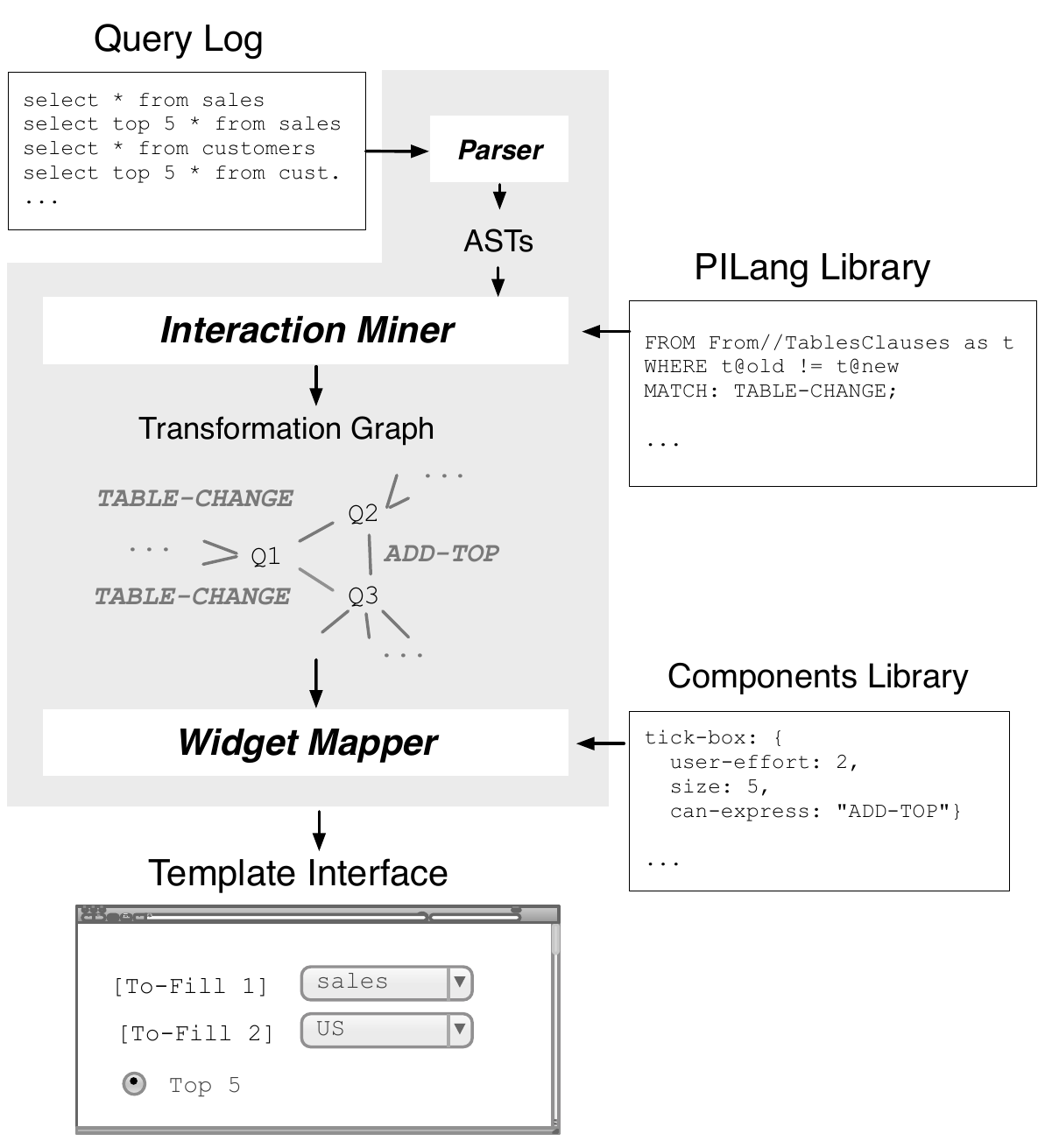}
 	\caption{Overview of PI's architecture.}
    \label{fig:architecture}
\end{figure}

\sys{} generates tailored interfaces from sequences of queries expressed in a query log.
It does so by mapping structural changes between the queries (e.g., adding an attribute to the \texttt{SELECT} clause) to user interactions in a generated web application (e.g., clicking a button, dragging a label). As illustrated in Figure~\ref{fig:architecture}, \sys employs a two step process. First, the {\it Interaction Miner} transforms the query log into a {\it transformation graph} where each query is a node and edges represent simple structural changes between the queries. Second, the {\it Interface Generator} maps the transformation graph to interactions in an application interface.

Detecting interactions in general is very challenging because differences between two queries may be arbitrarily complex. Our main insight is that the set of commonly used UI components and interactions---such as form elements, selection boxes, hovering, clicking---have a limited expressiveness. This observation simplifies the types of query differences that the system must consider. The rest of this section describes the key design consideration for each system component.

\stitle{Parsing.} Although this paper focuses on precision interfaces for SQL query logs, our vision is to build a general system that can be applied to different programming languages.
To this end, program strings are not the appropriate abstraction because they lack the necessary structure and semantics for detecting structural program changes.
Instead, we assume the existence of a language grammar and a parser that maps the program log into a sequence of abstract syntax trees (ASTs).

\stitle{Interaction Miner.}  This component runs tree matching algorithms between pairs of ASTs to identify sub-tree differences. One major challenge is to identify the types of differences that can be mapped to user interactions. For instance, mapping two completely different queries such as Q1 and Q3 in Figure~\ref{fig:query-pairs} may not help us build a practical interfaces. In our current implementation, developers use our \lang language to pre-specify the set of desirable tree differences and add them to the {\it Interactions Library} (Section~\ref{sec:language}). For instance, one \lang statement may be ``only a number in the \texttt{WHERE} clause changed'', or ``an expression was added to the \texttt{SELECT} clause''. If a statement matches a pair of ASTs, the Interaction Miner creates an edge in the transformation graph between the two corresponding queries that is typed by the \lang statement. Therefore, running tree matching between all pairs of queries in the query log will produce a transformation graph. In a near future, we will study automated methods to  map interactions and widget specifications.

\stitle{Interface Generator.} We take a two step approach towards generating interfaces. First we map edges in the transformation graph to abstract UI components (e.g., radio buttons, slider, hover) in the interface. This task is a challenge because we must take into consideration the resulting interface's {\it coverage}---meaning the set of queries that the interface can express---as well as its {\it complexity}---meaning the difficulty for users to understand the interface. For instance, a trivial interface would simply map each query in the log to a button that executes the query and presents the results. Such a UI would have a high coverage but also high complexity. Similarly, edges that represent a numerical value change (e.g., changing a threshold in the \texttt{WHERE} clause) could be represented as a slider, a set of radio buttons, or a textbox. Each options has its own trade-offs, depending on the range of options that the widget must cover, how much complexity the interface allows and how frequently the component is accessed. Section~\ref{sec:mapping} describes our approach for selecting UI components and balancing these trade-offs.

Once we have selected a set of abstract UI components, the second challenge is to populate the components with data (e.g., specify the minimum and maximum values of a slider), lay out the components, and render the interface.
Our current implementation simply renders the components is a grid-based web template and allows the user to populate, customize and reposition them. We are working on automating this step.

\stitle{Discussion.}  Our graph-based formulation provides considerable flexibility in the types of interfaces that we can generate by simply changing the subset of query log that \sys analyzes.
For instance, we might generate a fully expressive but complex interface by considering the complete query log.
In contrast, partitioning the log by analyst generates analyst-specific interfaces to each analyst, while partitioning by task can generate task-specific interfaces.

\section{\lang}
\label{sec:language}

\begin{figure}
	\begin{subfigure}[b]{\columnwidth}
	\scriptsize
	\begin{Verbatim}[commandchars=\\\{\},frame=single]
// PI_Lang statement 1:
FROM project//projectclause AS cols
WHERE cols@old subset cols@new AND |cols@old| = |cols@new|+1
MATCH ???
	\end{Verbatim}
	\begin{Verbatim}[commandchars=\\\{\}]
// Example of matched queries for statement 1:
SELECT \textcolor{Red}{region, revenue} FROM clients
SELECT \textcolor{Red}{revenue} FROM clients
	\end{Verbatim}
	\end{subfigure}
    \par\bigskip

	\begin{subfigure}[b]{\columnwidth}
	\scriptsize
	\begin{Verbatim}[commandchars=\\\{\},frame=single]
// PI_Lang statement 2:
FROM from//tableclause//tablename as T
WHERE T@old not equal T@new AND |T| = 1
MATCH ???
	\end{Verbatim}
	\begin{Verbatim}[commandchars=\\\{\}]
// Example of matched queries for statement 2:
SELECT * FROM \textcolor{Red}{Clients}
SELECT * FROM \textcolor{Red}{Regions}
	\end{Verbatim}
	\end{subfigure}
	\caption{Examples of PI-Lang statements and matched pairs of queries.}
	\label{fig:pilang-examples}
\end{figure}

We now describe \emph{PILang}, a domain-specific language to express structural differences between two ASTs $T_1$ and $T_2$.

A PILang statement is composed of three clauses. The \texttt{FROM} clause specifies where differences occur. It binds range variables to paths in the ASTs, using an XPath-like syntax. The semantics is that $T_1$ and $T_2$ are identical except for the sub-trees rooted at the matching nodes.

The \texttt{WHERE} clause is a boolean expression over the range variables that specifies how they may differ. The statement generates a match when the expression evaluates to true. The suffixes \texttt{@new} and \texttt{@old} can be appended to a range variable to reference the corresponding nodes in $T_1$ and $T_2$, respectively.  Finally, we support convenience expressions to help perform set comparisons between the two versions of the path expressions.  For instance, \texttt{T@old subset T@new} specifies that new nodes were inserted into \texttt{T}, whereas \texttt{|T| = 1} checks that there is only one matching node.

The \texttt{MATCH} clause labels the statement. In our implementation, we model the range variables as relational tables and translate \lang into SQL.  In the future, it can also expose the range variables that have changed so that their values can be dynamically bound to UI component state.

To illustrate, the following statement identifies pairs of queries with different string literals in an equality expression within their \texttt{WHERE} clause (Figure~\ref{fig:query-pairs}):
\begin{Verbatim}
FROM where//expr[op="="]//strliteral AS T
WHERE T@old not equal T@new AND |T| = 1
MATCH change_where_equal
\end{Verbatim}
The \texttt{FROM} clause matches all string literal nodes that are children of equality expressions in the filter clause.  These nodes are bound to \texttt{T}.
The \texttt{WHERE} clause checks that there is a single string literal that has changed (and implicitly that nothing else in the ASTs have changed).
If there is a match, then we add an edge between the two input ASTs and label the edge \texttt{change\_where\_equal}.

Figure~\ref{fig:pilang-examples} presents two additional examples of PI-Lang statements, along with matching pairs of queries.
Note that \lang statements are language agnostic and can be expressed over any programs that can be parsed into ASTs.
Currently, we are developing a library of standard transformations for SQL and plan to extend support for both query languages such as SPARQL and HIVE, as well as programming languages such as R and Python.
\section{Interface Generation}
\label{sec:mapping}
\begin{figure}
	\centering
	\includegraphics[width=.8\columnwidth]{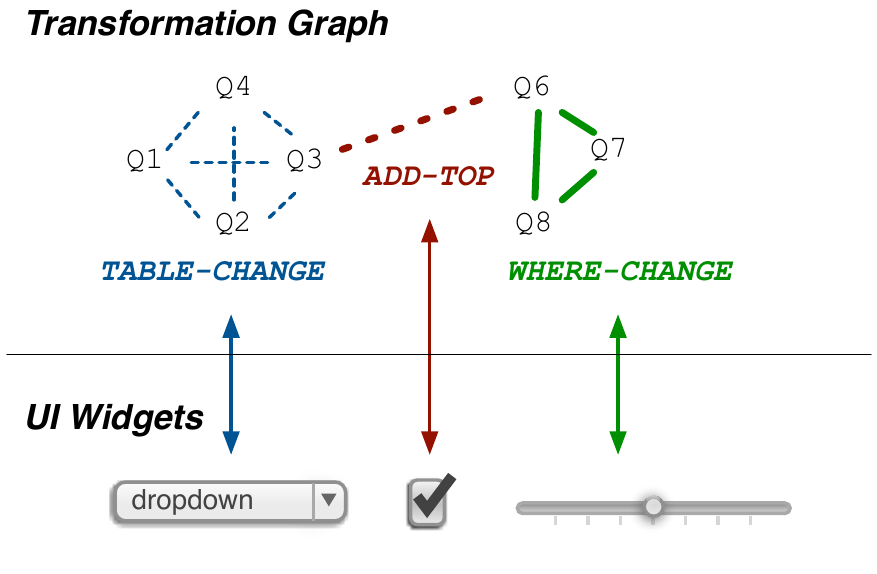}
  \caption{Examples of mapping the transformation graph to UI components.}
    \label{fig:mapping}
\end{figure}

We model the interface generation problem as identifying a mapping from sets of edges in the transformation graph to UI components that can express those edges.    For instance, Figure~\ref{fig:mapping} shows how edges that describe change the table in the \texttt{FROM} clause of a query may be mapped to a dropdown to select from the set of tables in the database; adding a \texttt{TOP 5} clause may map to a check box, whereas changes to a numerical attribute may map to a textbox or a slider.

In general, there can be many possible mappings to generate interfaces, and the natural question is ``what is a good interface?''.  Interface theory literature has decomposed the data analysis process into high level steps and identified the sources of friction that can impede user progress~\cite{lam2008framework, norman1988psychology}.  These sources include mapping high level goals to interface operations---which is impeded by complex interfaces---and fatigue from physically performing the operations. Based on this theory, our optimization follows three principles: \emph{coverage}, \emph{simplicity}, and \emph{efficiency}.

The interface should maximize \emph{coverage} in terms of the proportion of the graph that the interface can express.  Trivially, an interface can achieve full coverage by mapping each program to a button that executes the corresponding program when pressed.
However, such an interface will have high complexity and it will be challenging for a user to identify the appropriate button to click.
For this reason, we emphasize interface \emph{simplicity} by reducing the set of interaction components that are used in the interface.
However, a large query input box has full coverage and is simple, but defeats the original purpose of designing an interactive interface.
Thus, we seek to maximize \emph{efficiency}, which is modeled as the amount of human effort needed to express any given analysis.

Given a transformation graph $(\mathcal{V}, \mathcal{E})$ with nodes $\mathcal{V}$ and edges $\mathcal{E}$, a mapping $M = \{(E_i, i_i) | E_i \subseteq  \mathcal{E}, i_i \in \mathcal{I} \}$ maps a subset of edges $E_i$ to an interface component $i_i$ selected from a pre-defined interaction library $\mathcal{I}$.   The overall problem statement is:

\begin{definition}[Component Mapping]\label{eq:prob}
  Given a transformation graph $(\mathcal{V}, \mathcal{E})$, identify the optimal mapping
  \begin{flalign}
    M^* = {\text{argmin}}_{M}\ C_e(M)\\ \nonumber
          s.t.\ \                 C_c(M) < S_{max}\label{eq:constraint}
  \end{flalign}
\end{definition}

We seek to minimize the interaction cost $C_e(M)$ to transform any query from the log to any other, subject to a constraint on the interface complexity $C_c(M)$.

$C_e(M)$ is the average cost to transform between all the queries in the log $q_i$ and $q_j$.  We assume that it costs $c_e(i_i)$ to traverse an edge in the graph by using a given interaction $i_i \in M$.  Thus, the cost to transform between the two queries $c_e(q_i, q_j; M)$ is the minimum cost path that only uses interactions in $M$.  If such a path  doesn't exist, then we assign a default cost $penalty$.  With those notations, we can express $C_e(M)$ as the average cost between all query pairs in the log:
\begin{equation}
  C_e(\mathcal{L}, M) = \frac{1}{\left| \mathcal{L} \right|}
    \sum_{q_i, q_{j} \in \mathcal{L}^2} \mathit{c_e}(q_i, q_j; M)
    \label{eq:objective}
\end{equation}
Our prototype simply considers all adjacent pairs of queries in the log.

We approximate the interface complexity by assigning each UI component a complexity score $c_c(i)$, and model the total interface complexity as the sum of all components:
$$C_c(M) = \sum_{(e,i) \in M} c_c(i)$$
In future versions, we intend to use complexity measures from the interface literature~\cite{parush1998evaluating,shneiderman2010designing}.
\begin{figure}[t]
  \scriptsize
  \begin{Verbatim}[commandchars=\\\{\},frame=single]
SELECT "ontime"."distance" AS "distance",
    SUM("ontime"."arrdelay") AS "sum:arrdelay:ok",
    SUM("ontime"."depdelay") AS "sum:depdelay:ok"
FROM "public"."ontime" "ontime"
GROUP BY 1
HAVING (MIN("ontime"."distance") >= 30.99)
   AND (MIN("ontime"."distance") <= \textcolor{Red}{4983.00}))

SELECT "ontime"."distance" AS "distance",
    SUM("ontime"."arrdelay") AS "sum:arrdelay:ok",
    SUM("ontime"."depdelay") AS "sum:depdelay:ok"
FROM "public"."ontime" "ontime"
GROUP BY 1
HAVING (MIN("ontime"."distance")  >= 30.99)
   AND (MIN("ontime"."distance") <= \textcolor{Red}{2863.00}))
  \end{Verbatim}
  \caption{Pair of queries from Tableau's logs, with a value change in the \texttt{WHERE} clause.}
  \label{fig:diffs-ontime}
\end{figure}

\stitle{Solution Sketch.}
The problem described in Definition~\ref{eq:prob} is NP-hard, as it is a generalization of the knapsack problem. We approximate the solution with a greedy heuristic. At each step, PI computes all the possible widget-transformation assignments, eliminates those that violate the complexity constraint, and choses the one which leads to the best improvement of the objective function $C_e(M)$. The system then removes all the edges and vertices concerned with the corresponding transformation, and reiterates the procedure on the reduced graph. The algorithm stops when there is no space left on the interface, that is, when $C_c(M) \geq S_{max}$ .

\section{Early Results}
\label{sec:experiments}

\begin{figure}[t]
  \centering
  \includegraphics[width=.8\columnwidth]{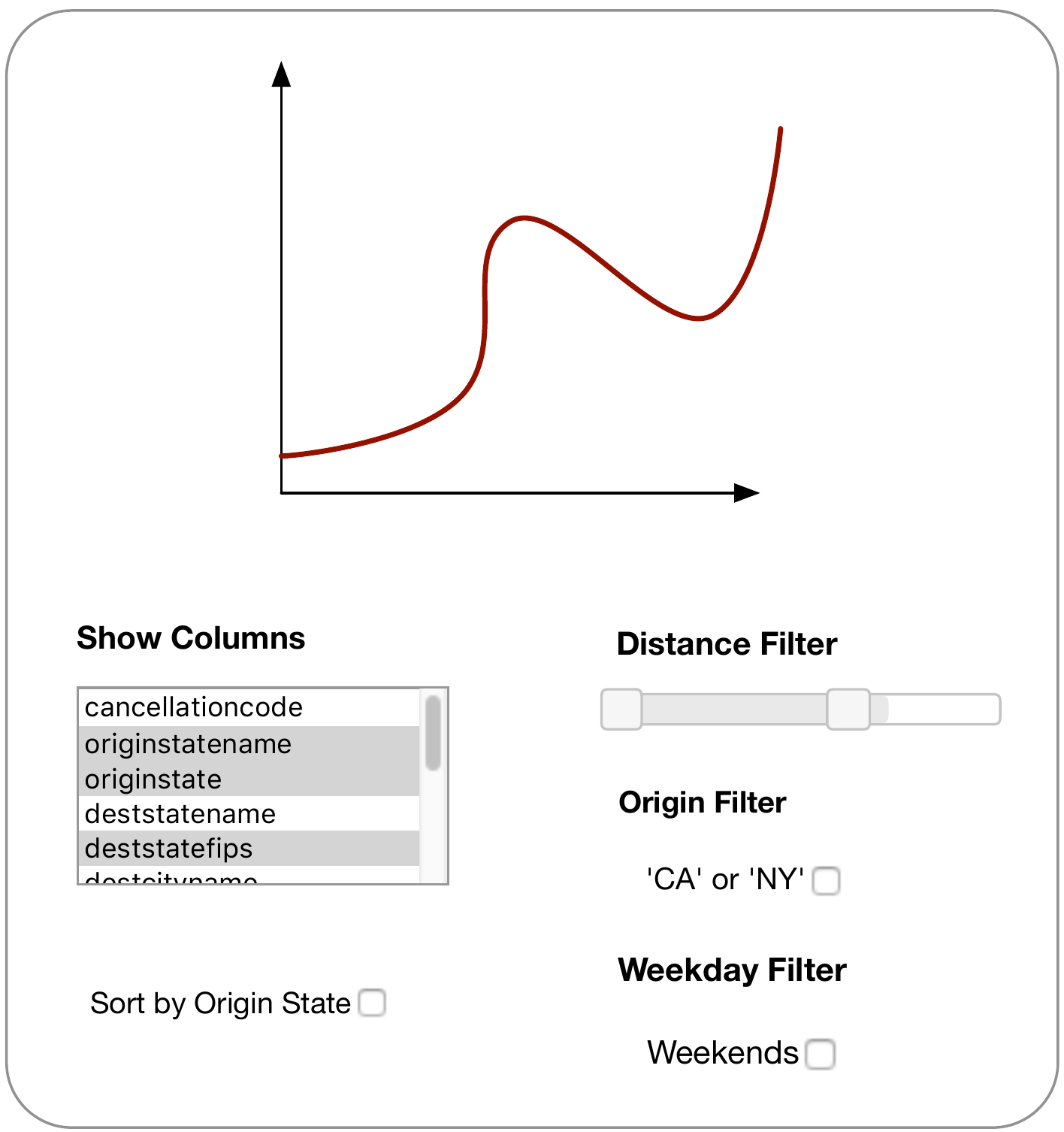}
  \caption{Interfaces generated by Precision Interfaces for the first student, with a mock-up output. We filled the data in the components, placed them on the page and wrote captions through the template generated by our system. }
  \label{fig:rendered1}
\end{figure}


 We now present experiments with on our prototype implementation. We asked Computer Science students to analyze the On-Time Database\footnote{521,000 rows and 91 columns. \texttt{https://www.transtats.bts.gov}} with Tableau and collected the generated SQL queries. Our aim is to show that (1)~each user only uses a small set of analysis operations, (2)~\sys{} can recognize those patterns from the query logs and (3)~\sys{} can automatically produce custom interfaces for each user.

\stitle{Setup.}  We asked students to answer 3 out of 12 predetermined questions (e.g., ``how delayed are flights to California'') and answer one free-form question (``tell us something you find interesting''). We report an analysis based on the two longest query logs we collected (from two different students), which contain  167 and 137 queries respectively.
We used 9 \lang{} statements and the default generation parameters for both logs, and simply report our results.
\begin{figure*}[t!]
  \centering
  \includegraphics[width=1.3\columnwidth]{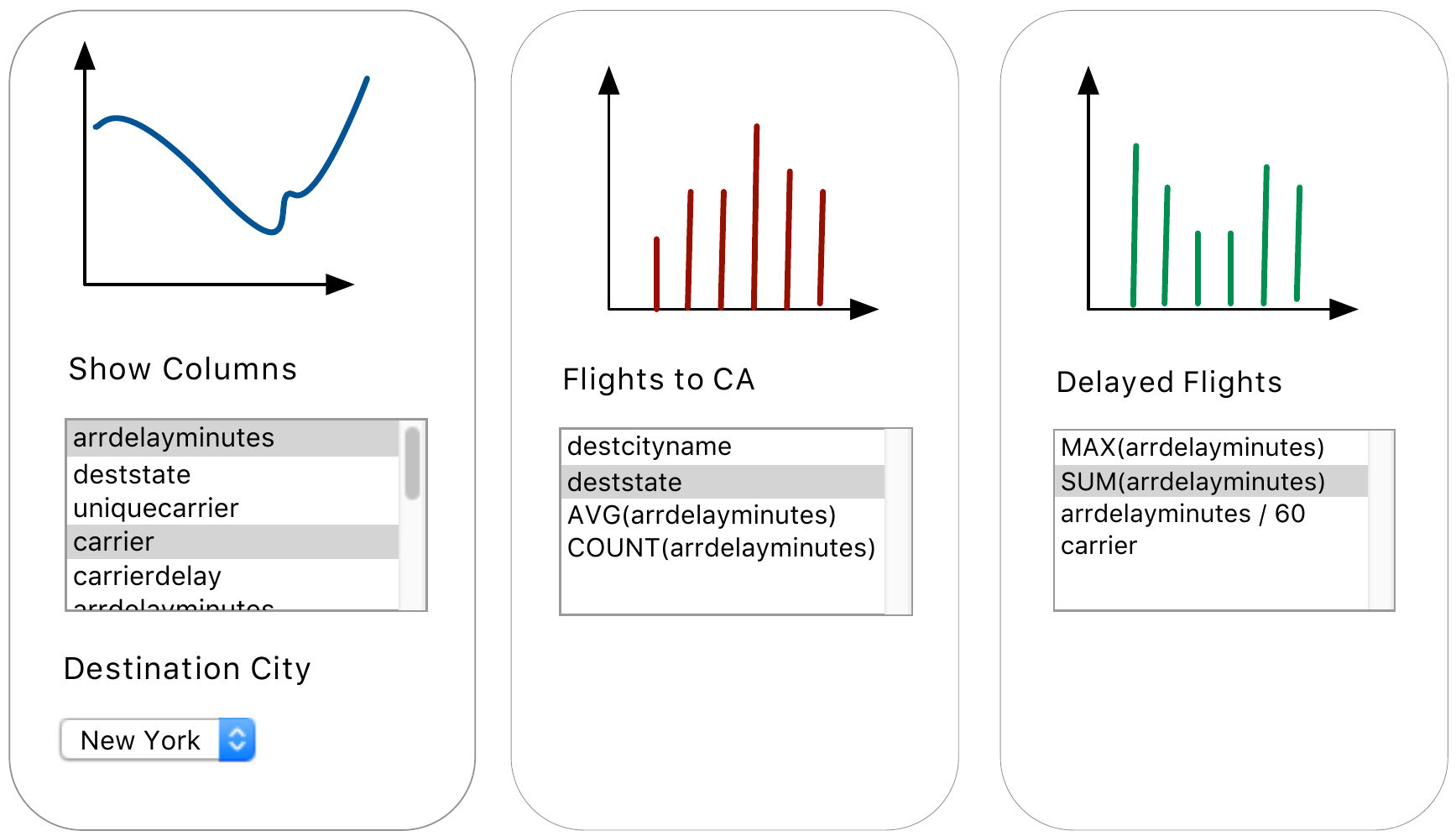}
  \caption{Interfaces generated by Precision Interfaces for the second student, with mock-up outputs.}
  \label{fig:rendered2}
\end{figure*}

\stitle{Results.} Figure~\ref{fig:rendered1} demonstrates the first interface generated by our system, along with mock-up outputs\footnote{This paper focuses on UI inputs. See related work~\cite{idreos2015overview, mackinlay2007show, wongsuphasawat2016voyager} for automatic visualization generation.}. Our first student decided to analyze the cause of flight delays by projecting and selecting subsets of the OnTime dataset. The interface presented Figure~\ref{fig:rendered1} expresses 166 out of the 167 queries that she produced, using only 5 components.

In this interface, the main component is the ``Show Columns'' list-box on the top left, which lets the student select which columns  of the table to visualize. The tick box at the bottom toggles sorting by State. The right part of the interface consists of three filters. The top filter restricts the flight distance using a range slider. The second one toggles whether or not to filter the flights from either New York or California. The bottom filter restricts the analysis to weekend flights.

Figure~\ref{fig:rendered2} represents the UI generated for the second student. This interface is more complex because the user performed three distinct subtasks. The leftmost panel lets her analyze all the flights in the database. The central panel focuses on flights to California. The rightmost panel focuses on delayed flights. The interface covers 120 out of 137 queries in the log.

\begin{figure}[t]
	\centering
	\includegraphics[width=1\columnwidth]{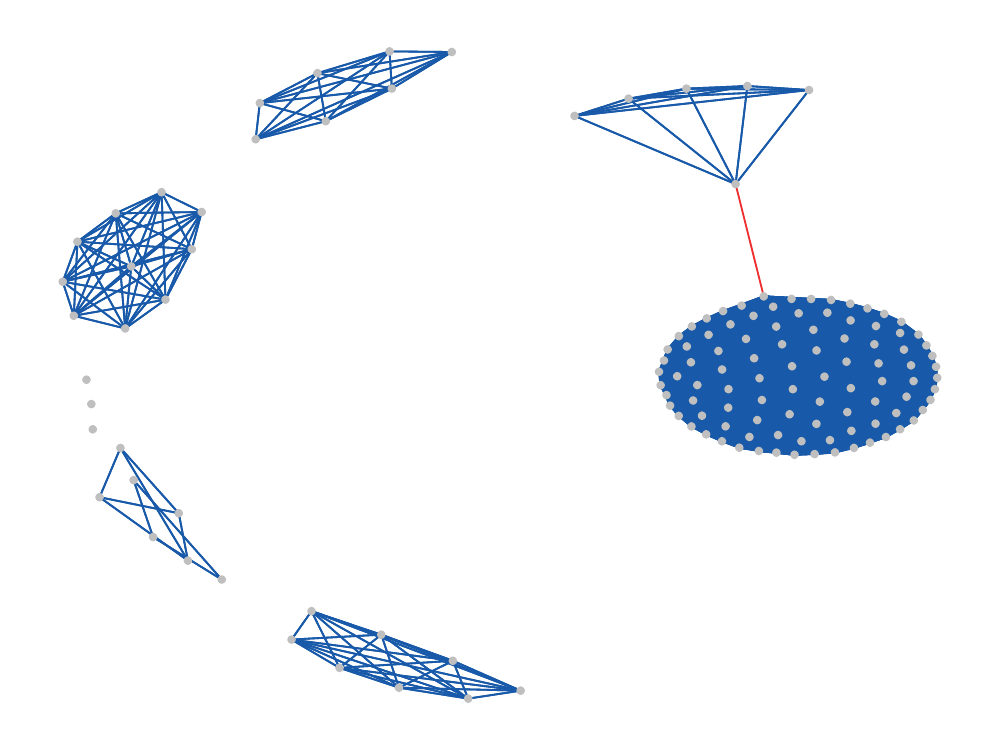}
 	\caption{Transformation graph for one of the analysis sessions. Blue edges describe changes in the \texttt{SELECT} clause, red edges describe changes in the \texttt{WHERE} clause.}
    \label{fig:graph}
\end{figure}

To understand why \sys{} chose to generate three separate interfaces, we plot the transformation graph in Figure~\ref{fig:graph}. Recall that each edge corresponds to a transformation between two queries---for instance, the blue edges represent changes in the \texttt{SELECT} clause. Thus each isolated cluster represents a distinct set of analyses, either by focusing on a different subset of the database, or by executing structurally different queries. If the user had performed incremental changes between the clusters, \sys{} would have created a single interface to express them all.

\if{0}
\subsection{Related Work}
\label{sec:related}

\stitle{Development Framworks.} Researchers and companies have presented several  development tools for non technical users, such as Mac Automator, Scratch~\cite{resnick2009scratch}, Sikuli~\cite{yeh2009sikuli} or even Microsoft Office's VBA. Those tools are efficient and provably user-friendly, but they still require learning a language, coding and debugging. Our aim is to automate the process further.

\stitle{Interface Generation.} Authors have been trying to automate interface design for several decades. We refer readers to Nichols and Faulring's 2000 survey for an overview of searly methods~\cite{myers2000past}. The most relevant systems are \emph{model-based}, i.e., users guide the user interface generation process through formal specifications. Early examples are UIDE~\cite{kim1993providing} and Jade~\cite{zanden1990automatic}, both from the early 90's. According Nichols and Faulring, those have failed to reach mainstream because of their unpredictability, and because their users need to learn new languages. We acknowledge the lesson. To cope with unpredictability, we will investigate how to include the user deeper in the generation process. To cope with the new language, we will develop ready-made libraries of PI-Lang statements for popular languages (e.g., SQL, R, Python) and investigate how to detect interactions automatically.

More recently, authors have tackled interface generation for more specific use cases. Nichols et al.~\cite{nichols2004improving, nichols2006uniform} have focused on appliances such as alarm clocks, remote controls or media players. Weld et al. have published the summary of a decade's work on UI personalization~\cite{weld2003automatically}. Among other systems, SUPPLE helps adapting interfaces for motor-impired users~\cite{gajos2010automatically}. To the best of our knowledge, none of those systems use program logs.
\fi

\section{Conclusions and Future Work}
\label{sec:conclusion}

We have argued for {\it Precision Interfaces} and described our prototype system that generates such interfaces from program logs.  We described a domain specific language for specifying interesting structural changes between program parse trees, modeled the program log as an \emph{interaction graph}, and described a graph-based algorithm for mapping the graph to a set of interface components.   Our case study on query traces generated from several open-ended Tableau exploration sessions showed that different users (and even the same user) perform different types of analysis tasks, and \sys generated simple, custom interfaces for each task.   This research is still in the early stages, and we are actively working on the following extensions to the system.

\stitle{Optimizations.}
In practice, interaction graphs are extremely dense because most transformations are transitive.
Consider changing the table name in the \texttt{FROM} clasue.  If $Q_1$ can transform into $Q_2$, and $Q_2$ can transform into $Q_3$, then $Q_1$ certainly transforms into $Q_3$.
Similarly, many transformations are also reflexive.
This forms dense, strongly connected clusters in the graph with $O(N^2)$ edges for $N$ queries.

We are exploring blocking-based techniques~\cite{baxter2003comparison} that can avoid all pair-wise comparisons within a dense cluster of programs, as well as sampling techniques that can guarantee that the sampled interaction graph will result in equivalent generated interfaces.

\stitle{Rendering.} This paper described generating interface components that the user can use to express program changes, however we have actively not considered how program {\it outputs} should be rendered in the interface.   A simple approach is to provide default tabular visualizations or use existing visualization generation techniques~\cite{mackinlay1986automating,mackinlay2007show,wongsuphasawat2016voyager}, however we are also exploring ways to identify the rendering functions in the programs themselves and incorporate them into the interface.

\stitle{Incremental Maintenance.} We envision running \sys as a system process that monitors and recommends new interfaces automatically.  In such a setting, the program log is constantly evolving and it is desirable to generate new or enhanced interfaces without re-running the whole pipeline. Similarly, it is desirable to identify and discard obsolete interfaces.  We are exploring incremental approaches to dynamically maintaining the interaction graph~\cite{mondal2012managing} as well as the set of generated interfaces.

\textbf{Automatic \lang.} The quality of the generated interfaces depends on a rich set of \lang statements that represent the core set of structural changesin the log.  Identifying and specifying these statements is a key challenge.  We are working on automatically inferring \lang statements from program logs and richer interface component specifications.  For instance, consider a simple slider---it is parameterized by the minimum and maximum numbers, and can modify a single number.  This specification naturally restricts the classes of \lang statements that it can map to.  Similarly, we might not consider complex strutural changes such as adding and removing quantification expressions because the only interface components that may express those are text boxes or specially crafted interface components.

\stitle{Acknowledgements: } We thank Yifan Wu for the initial inspiration, Anant Bhardwaj for data collection, Laura Rettig on early formulations of the problem, and the support of NSF 1527765 and 1564049.

\small
\balance
\bibliographystyle{abbrv}
\bibliography{hilda}

\end{document}